\documentclass[twocolumn]{aastex62}
\usepackage{verbatim}
\received{January 1, 2018}
\revised{January 7, 2018}
\accepted{\today}
\submitjournal{ApJ}

\shorttitle{UV plane of AGN}
\shortauthors{{\'S}niegowska et al.}

\begin{document}

\title{Quasar main sequence in the UV plane}

\correspondingauthor{Marzena {\'S}niegowska}
\email{marzena.sniegowska@student.uw.edu.pl}

\author[0000-0002-0786-7307]{Marzena {\'S}niegowska}
\affiliation{Center for Theoretical Physics, Polish Academy of Sciences, Al. Lotnik{\'o}w 32/46, 02-668 Warsaw, Poland}
\affiliation{Warsaw University Observatory, Al.  Ujazdowskie 4, 00-478 Warsaw, Poland}
\affiliation{Nicolaus Copernicus Astronomical Center, Polish Academy of Sciences, Bartycka 18, 00-716 Warsaw, Poland}


\author[0000-0003-4084-880X]{Szymon Koz{\l}owski}
\affiliation{Warsaw University Observatory, Al.  Ujazdowskie 4, 00-478 Warsaw, Poland}

\author[0000-0001-5848-4333]{Bo{\.z}ena Czerny}
\affiliation{Center for Theoretical Physics, Polish Academy of Sciences, Al. Lotnik{\'o}w 32/46, 02-668 Warsaw, Poland}

\author[0000-0002-5854-7426]{Swayamtrupta Panda}
\affiliation{Center for Theoretical Physics, Polish Academy of Sciences, Al. Lotnik{\'o}w 32/46, 02-668 Warsaw, Poland}
\affiliation{Nicolaus Copernicus Astronomical Center, Polish Academy of Sciences, Bartycka 18, 00-716 Warsaw, Poland}

\author[0000-0002-2005-9136]{Krzysztof Hryniewicz}
\affiliation{National Centre for Nuclear Research, Pasteura 7, 02-093 Warsaw, Poland}
\affiliation{Nicolaus Copernicus Astronomical Center, Polish Academy of Sciences, Bartycka 18, 00-716 Warsaw, Poland}



\begin{abstract}

Active galaxies form a clear pattern in the optical plane showing the correlation between the Full Width at Half Maximum (FWHM) of the H$\beta$ line 
and the ratio of the Equivalent Width (EW) of the optical FeII emission and the broad EW(H$\beta$). This pattern is frequently referred to as the 
quasar main sequence. In this paper, we study the UV plane showing the FWHM of MgII line against the ratio of the EW of UV FeII emission to the broad EW(MgII). 
We show that the UV plane trends are different, with the underlying strong correlation between the FWHM(MgII) and the EW(MgII). 
This correlation is entirely driven by the choice of the continuum used to measure the EW(MgII). 
If instead of the observationally determined continuum, we use a theoretically motivated power-law extrapolated from the wide wavelength range, the behaviour of the 
FWHM vs. EW for MgII becomes similar to the behaviour for H$\beta$. Such a similarity is expected since both the lines belong to the Low Ionization group 
of emission lines and come from a similar region. We discuss the behaviour of the lines in the context of the Broad Line Region model based on the presence of dust in the accretion disk atmosphere.

\end{abstract}

\keywords{quasars: emission lines, catalogs, methods: statistical}


\section{Introduction} \label{sec:intro}

Active galactic nuclei (AGN) are complex sources. A variety of their observed properties have led to various classification schemes
\citep[for a recent review, see e.g.][]{pad17}.
Even if we limit ourselves to the radio-quiet unobscured AGN with dusty, molecular torus not crossing our line of sight towards the
nucleus, a considerable complexity shows up in the broad band spectra and the emission line properties. It is not surprising since,
from the theoretical point of view, we expect that an active nucleus should be parameterized by the mass of the central
supermassive black hole, its spin, the accretion rate, and the viewing angle. This last quantity is likely to be important even in
unobscured objects since the key elements of an active nucleus: the accretion disk and the Broad Line Region (BLR) are considerably
flattened.

Nevertheless, the Principal Component Analysis (PCA) of \citet{bg92}  showed 
that by combining 13 measured AGN spectral properties 
it was possible to construct an Eigenvector 1 (EV1) relation which was responsible for a significant fraction (29.2 \%)  of the dispersion present in their sample.
Thus AGN also form a main sequence, although the correlations are by no means as tight as in the case of stellar main sequence.

The key
quantity in EV1 was the parameter $R_{Fe_{opt}}$, which is the ratio of the equivalent width (EW) of FeII lines in the optical band (4434-4684 \AA) to the EW of broad H$\beta$ line. Later
on, many studies specifically concentrated on the optical plane of EV1 made by $R_{Fe_{opt}}$ and Full Width at Half Maximum (FWHM) of broad H$\beta$ line
\citep[e.g.][]{sul00,sul02,zam10,marziani2018,2019ApJ...882...79P}. In parallel, full multi-parameter PCA approach also continued \citep[e.g.][]{kur09,marziani2018}, 
and a spectral PCA approach started by \citet{francis92} was also further developed \citep{roc17,dav18}. However, from the point of view of theoretical interpretation, simpler studies limited to a single plane are particularly attractive.

The existing correlations lead to a characteristic shape covered by the data points in the optical plane. This shape does not always look identical since various authors apply different criteria in selecting objects for plots to enhance the pattern. For example, \citet{sh14} use a smoothing box of the size 0.2 in $R_{Fe_{opt}}$ and of 1000 km s$^{-1}$ in FWHM, and plot objects only if there are more than
50 objects in the smoothing box. On the other hand, \citet{zamfir2008} have a relatively small number of objects when they introduce their division of the optical plane into several sub-classes, and objects are included in the plot if they belong to these specific sub-classes \citep[e.g.][]{marziani2018,2019ApJ...882...79P,2020CoSka..50..293P}.

In this paper, we study the UV plane in a close analogy to the optical plane of EV1, by replacing the optical FeII emission with
the UV FeII emission, and the H$\beta$ line with the MgII line.
In Section~\ref{sec:method} we describe the data we use in the current work.
The results for the optical and the UV planes are given in Section ~\ref{sec:results}. Section~\ref{sec:results} also includes the correlations of the line widths and the line EWs along with the respective Fe II pseudo-continua.
In Section~\ref{sec:discussion}, we discuss the expected properties of the BLR at the basis of the assumed BLR model. We also show how the observed correlation between the line width and the line EW for Mg II changes if, instead of a continuum underneath the line we measure the line EW using a theoretically motivated asymptotic power-law. We summarize our results from this work in  Section~\ref{sec:conclusions}.

\section{Method}
\label{sec:method}

We select a subsample of quasars with the parameters determined by the QSFit \footnote{Quasar Spectral Fitting package. We use the parameters determined using their version 1.2.4 available through \href{http://qsfit.inaf.it/}{http://qsfit.inaf.it/})} \citep{calder17}, which contains 71 261 objects.
We did not use the software ourselves but downloaded the QSFit catalog prepared by the authors using the spectroscopic measurements from the Sloan Digital Sky Survey (SDSS). The fitting method of \citet{calder17} assumed a power-law model for the AGN continuum, a Balmer continuum component, and a host galaxy contamination based on an elliptical galaxy template. The H$\beta$ and MgII line profiles were measured assuming two Gaussians 
(one for the narrow component and the other for the broad component), and FeII pseudo-continuum in the optical and UV bands were fitted using the appropriate templates.
Additional Gaussian components were also added whenever necessary. 

Our subsample contains all quasars from this catalog which have both MgII and H$\beta$ lines within the observed range. We limited ourselves to sources with the measurement errors of each of the following quantities smaller than 20\%. The fraction of objects that satisfy each criterion are marked within parenthesis:
FWHM(H$\beta$) (15 179/71 261), FWHM(MgII) (9215/71 261), EW(MgII) (9155/71 261), EW(H$\beta$) (8947/71 261), EW(FeII$_{opt}$) (6861/71 261), and EW(FeII$_{UV}$) (3536/71 261). Next we add two more selection criteria. We request the EW(OIII5007) measurement to be non-zero, which reduces the sample to 2967 objects, and we exclude for which the continuum slopes (CONT4\_SLOPE) at 4210 \AA\ were not measured/reported.

\begin{figure}
\plotone{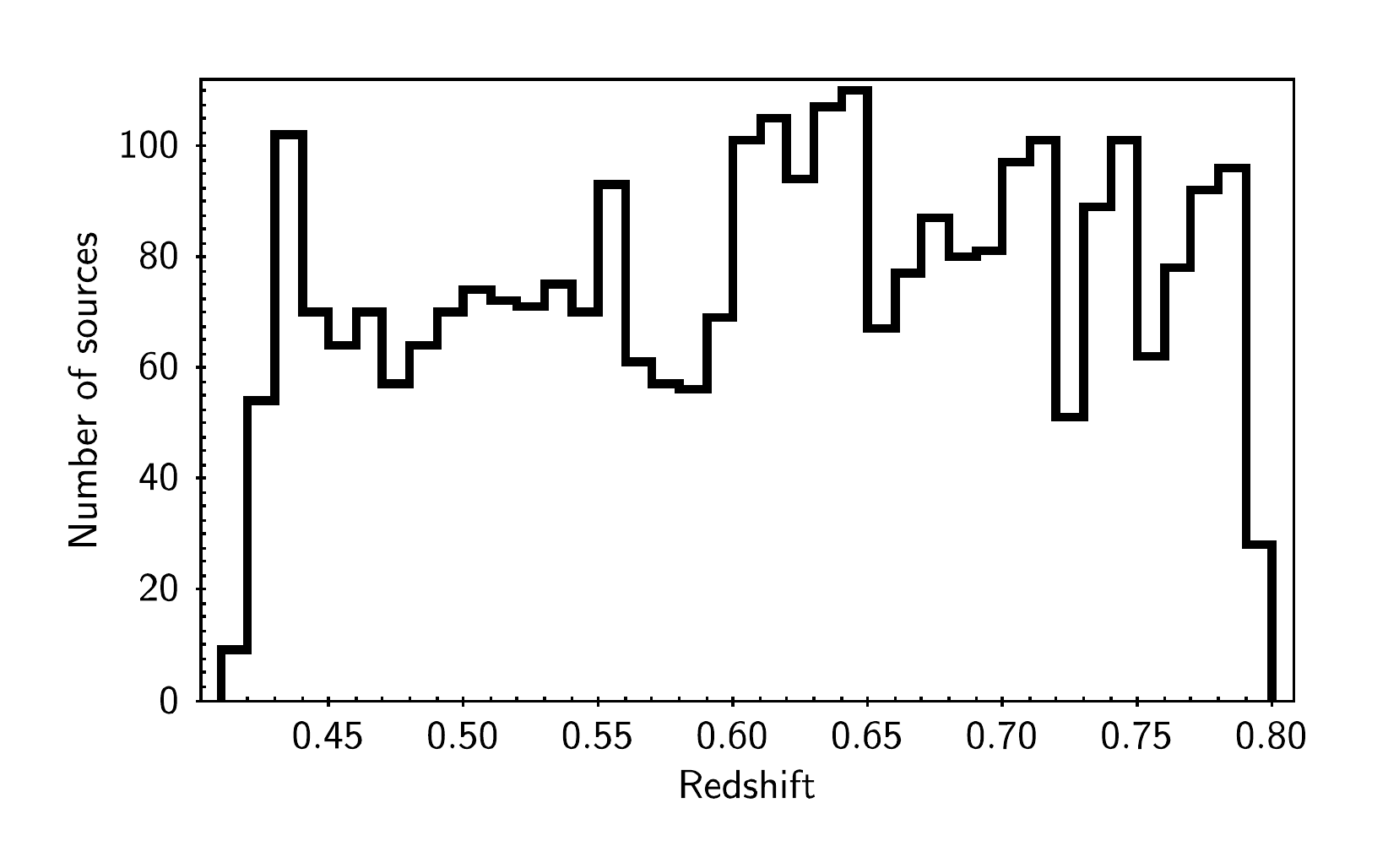}
\caption{The redshift distribution of the studied quasar sample.}
\label{fig:redshift}
\end{figure}

\begin{figure}
 \plotone{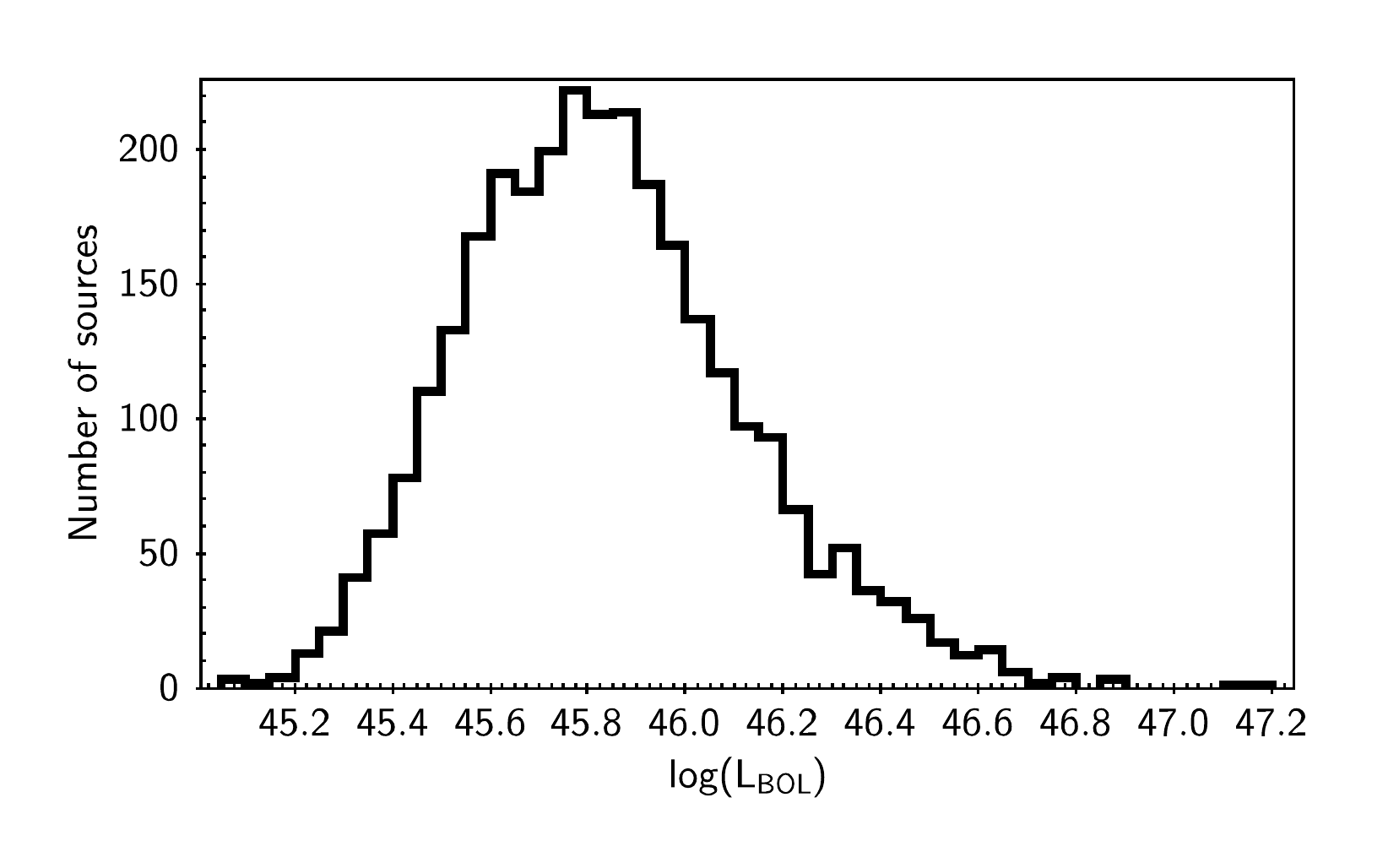}
  \caption{The distribution of the bolometric luminosity in the studied quasar sample.} \label{fig:lbol}
  \end{figure}

The total number of objects in the selected sample is 2962.
The sources cover the redshift range from 0.4 to 0.8 rather uniformly. 
(see Figure~\ref{fig:redshift}). All selected quasars are thus at intermediate redshifts due to the requirement of H$\beta$ coverage. Selected sources are moderately bright, the distribution of the bolometric luminosity in the sample is shown in Figure~\ref{fig:lbol}. 
The luminosities concentrate around values of the order of log(L$_{BOL}$ 45.8), typical for intermediate redshift quasars \citep{2011shang}. Sources are generally radio-quiet, and only a few radio-loud quasars are in the sample (237/2962). Quantitatively, the exact numbers for radio-loud and radio-quiet sources are difficult to give since for most sources the loudness parameter is not reported in the original \citet{s11} catalog. 
The median for the radio-loudness parameter (defined as the ratio of radio to optical luminosity) for the 237 objects is equal to 40.1.
The QSFit catalog does not provide this information.
The quality of the spectra in the sample is satisfactory, with the median S/N  per pixel for the rest frame 4750-4950 \AA ~region at 13.9, and the median S/N per pixel for the rest frame 2700-2900 \AA ~region at 15.3. 
 The typical relative measurement errors in the selected sample are reported as follows: for FWHM(H$\beta$)= 0.0603, FWHM(MgII)= 0.0439, EW(H$\beta$)= 0.0578, EW(MgII)= 0.0359, EW(Fe$_{opt}$)=  0.0838, and, EW(Fe$_{UV}$)=  0.0837.

The EW(FeII$_{opt}$) in the QSFit catalog is measured in the wavelength range 3500 - 7200 \AA\ (rest frame)  \citep{veron04}, and the
EW(FeII$_{UV}$) is measured in the
range 1250 - 3090 \AA~(rest frame) \citep{2001ApJS..134....1V}. The optical range for the quoted FeII emission is systematically different from the values given 
by \citet{s11} and \citet{sh14} who have used only the range from 4434~\AA\ to 4684 \AA, based on the \citet{1992ApJS...80..109B} prescription.
For each object we recalculate the EW(FeII$_{opt}$) from 3500 - 7200 \AA\ (rest frame) to 4434~\AA\ - 4684 \AA. We reconstruct the continuum for each object using the continuum luminosity and the spectral slopes computed and provided in the QSFit catalog. Smearing the iron template \citep{veron04} to v = 3000 km s$^{-1}$ for the broad component and v = 500 km s$^{-1}$ for the narrow component, and re-normalizing the flux, we compute the EW(FeII$_{opt}$)) for the new spectral window. This complex procedure is roughly equivalent to rescaling the EW(FeII$_{opt}$) values obtained from the QSFit catalog by a factor of $\sim 6$ for the majority of the sample objects. 




\section{Results}
\label{sec:results}

\begin{figure}
\plotone{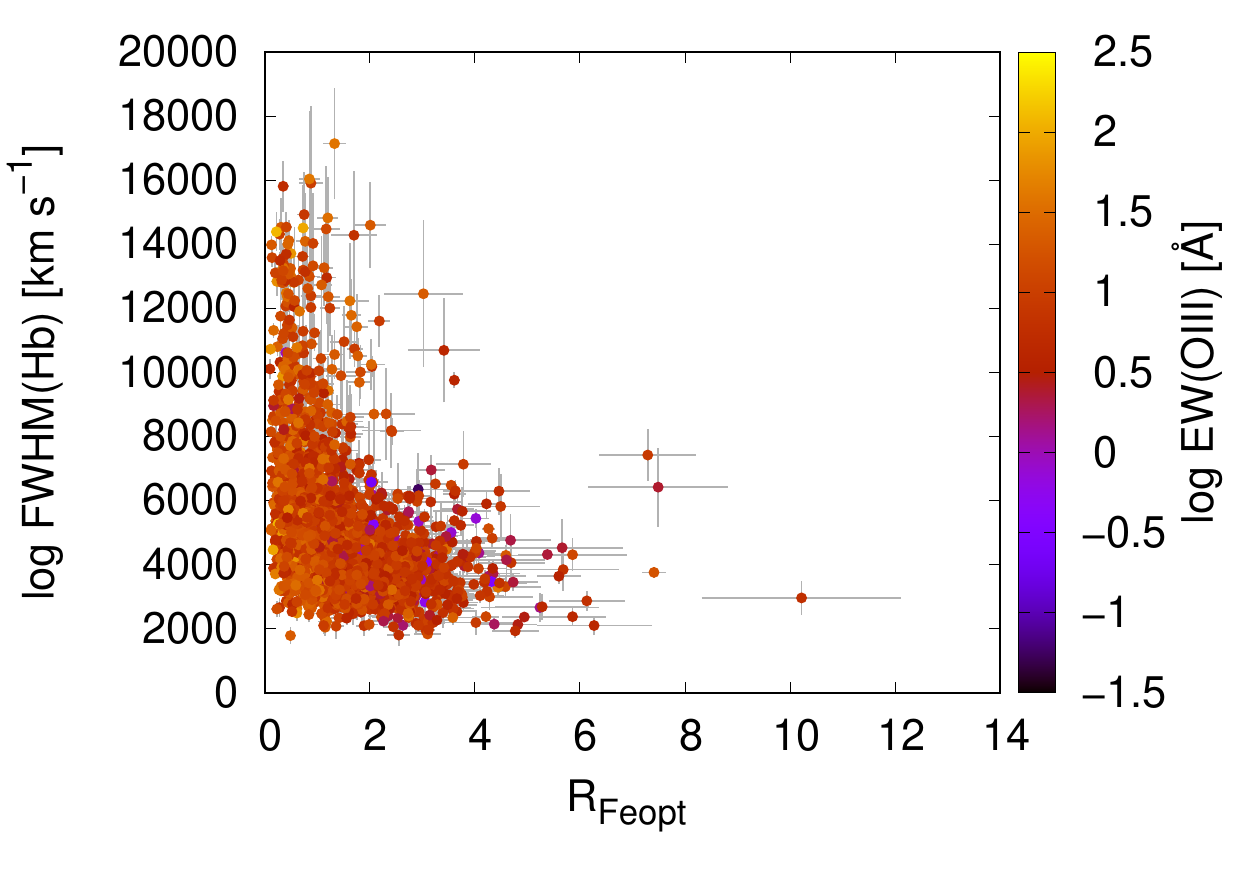}
\plotone{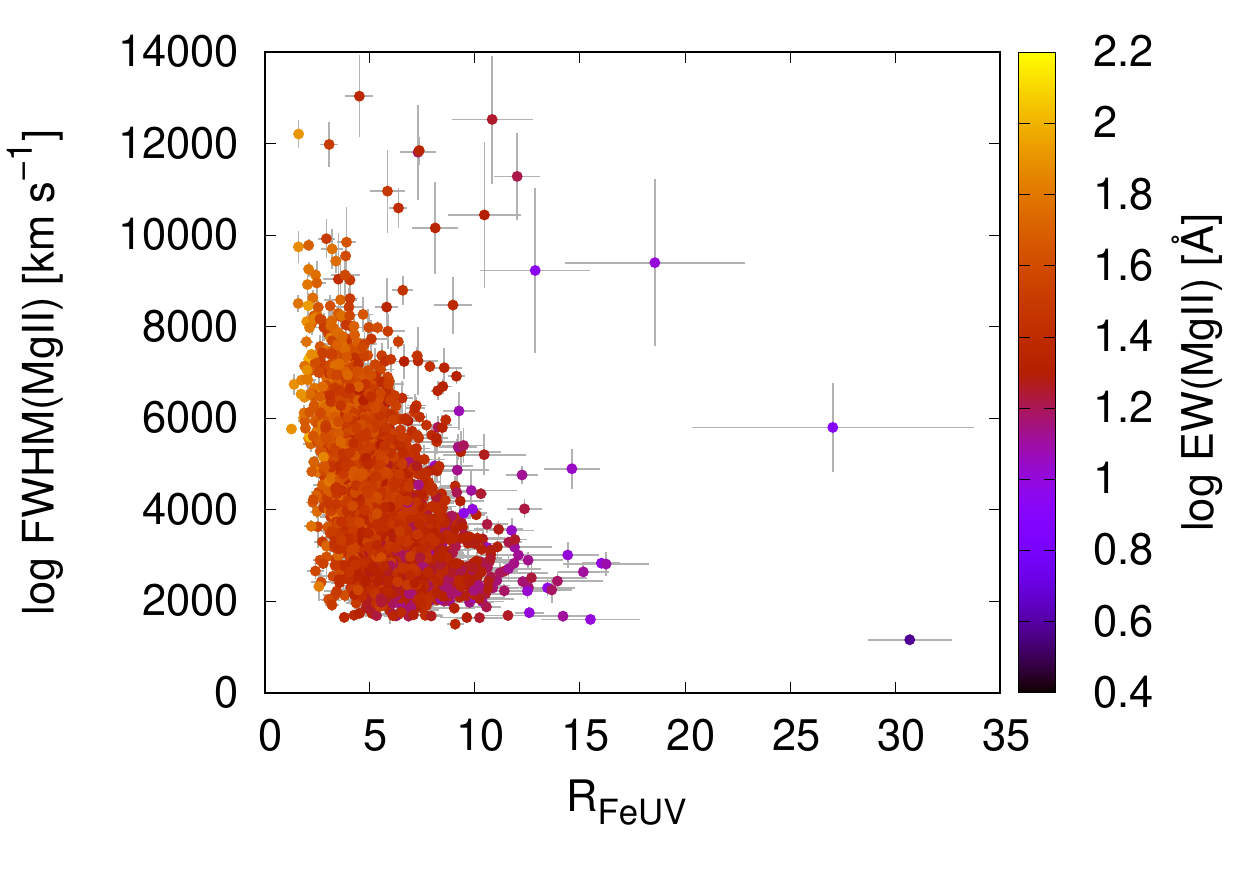}
  \caption{The optical plane (top) and the UV plane (bottom) for the subsample of quasars from the QSFit catalog with errors for each value. }
  \label{fig:planes}
  \end{figure}

For this subsample we construct the optical and the UV plane by plotting the FWHM(H$\beta$) vs. the ratio $R_{Fe_{opt}} = EW(FeII_{opt}) / EW(H\beta)$, and the FWHM(MgII) vs. the ratio \\
 $R_{Fe_{UV}} = EW(FeII_{UV}) / EW(MgII)$ (see Figure~\ref{fig:planes}) and compare the properties of the two planes. 
The optical quasar main sequence (upper panel of Figure~\ref{fig:planes}) is well visible as in the diagrams presented in the literature \citep{sul00,zam10,sh14,marziani2018}. 
It constitutes of a general trend of decreasing FWHM(H$\beta$) with increasing strength of the FeII$_{opt}$ contribution measured with respect to the H$\beta$.
The extension of the plot in $R_{Fe_{opt}}$ is larger (we have values above 2.25) than the values shown by \citet{sh14} since they 
limited their sample to objects densely populating the diagram while we did the selection at the basis of the fit quality.
We colored the objects with respect to the EW(OIII) and we see a general trend of these values decreasing with the $R_{Fe_{opt}}$.

The UV plane looks topologically similar (lower panel of Figure~\ref{fig:planes}). The range of the FWHM(MgII) is slightly narrower, but the same general trend of decreasing line width, 
this time with the $R_{Fe_{UV}}$ is visible. The range of the $R_{Fe_{UV}}$ values is quite different but this is due to the adopted wavelength range 
involved in calculating the EW($Fe_{UV}$) which is very broad. The points colored with the EW(MgII) show the same pattern
as in the optical plane, i.e. with respect to the EW([OIII]).

Therefore, the UV plane seems equivalent to the optical plane at a first glance. In principle, this is not surprising since the MgII line belongs
to the Low Ionization Lines (LILs) together with the H$\beta$ \citep{collin88}.  However, the connection between the optical and the UV emission of FeII 
is not so clear \citep{kovacevic15}. Therefore, we analyzed separately the dependence of the line width on the EW of the studied lines (H$\beta$ and MgII) and on the corresponding FeII emissions.

\begin{figure}
  \plotone{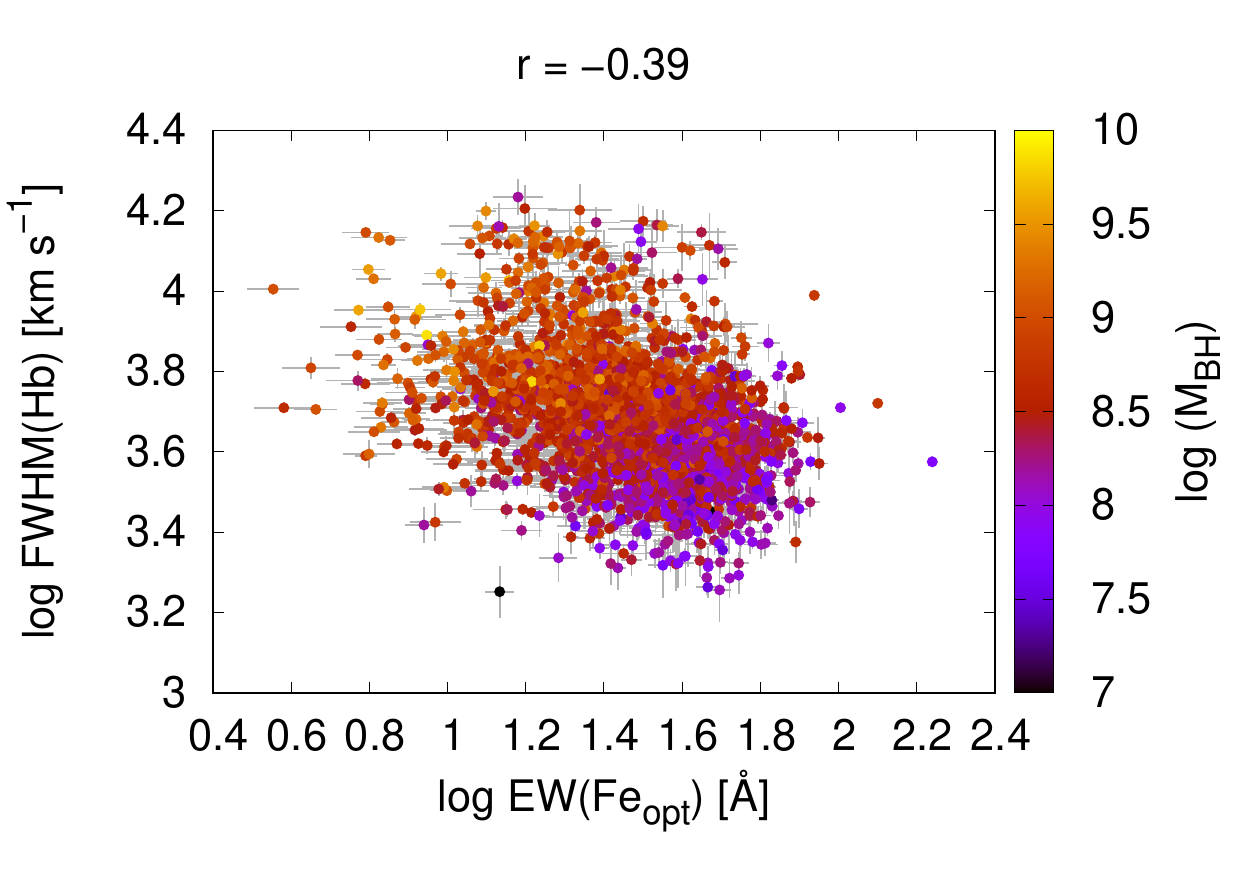}
  \plotone{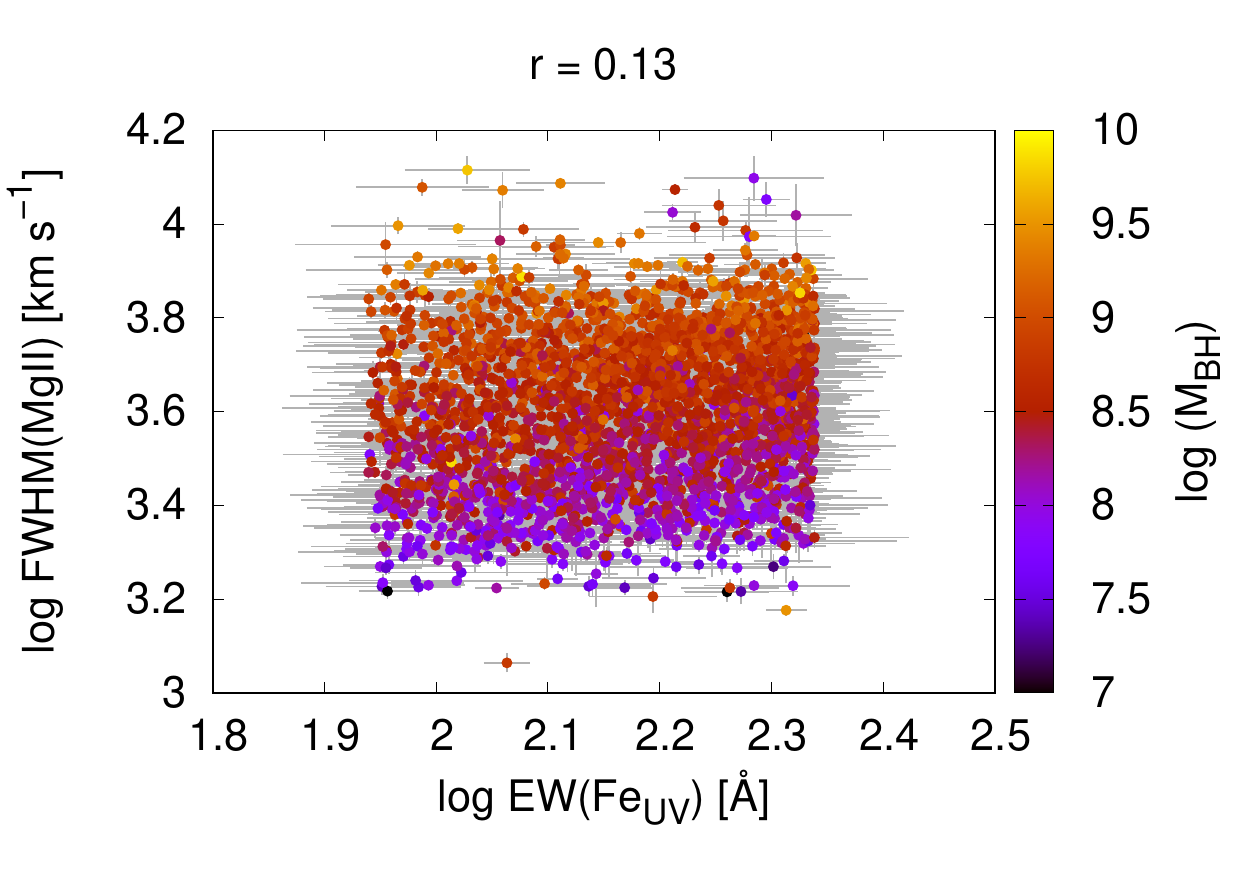} 
  \caption{The dependence of the FWHM(H$\beta$) on the EW(FeII$_{opt}$) (top), and the FWHM(MgII) on the EW(FeII$_{UV}$) (bottom) for the subsample of quasars from the QSFit. }
  \label{fig:FWHM_Fe}
  \end{figure}

The dependence between the FWHM of the line and the equivalent width of the corresponding line in the optical and in the
UV plane are shown in Figure~\ref{fig:FWHM_Fe}. To see more clearly the possible character of the dependence, we use logarithmic 
values instead of the traditional linear values reserved for the plots in the optical plane.
We can do that since we selected only high quality results
from the QSFit and the errors for small values of the FeII intensity do not distort the plot. 

In order to study the presented relation quantitatively, we calculate the Pearson correlation coefficient, r, and the correlation significance measured through the p-value which tests the null hypothesis that the coefficient is equal to zero (no effect). If P $>$ 0.05 the correlation is not significant, and if P $<$ 0.0027 the correlation is significant at more than 3 sigma level. We then use the following approach: if $|r| >$ 0.5, we determine the best-fit linear trend using the orthogonal regression fits, taking into account individual errors in each measurement. The method we apply is called the orthogonal distance regression (ODR) \citep{Boggs1989OrthogonalDR}. We use this approach to the data of  Figure~\ref{fig:FWHM_Fe} and in the figures henceforth. However, when r is below 0.5, then only a r$^2$ fraction ($<$ 0.25, or 25 \%) of the variance is due to correlated changes in the two studied variables. In such case the predictive power of the linear best fit is very limited, since the corresponding prediction interval is broad. Thus we show the best fits only for the limiting values $|r| >$ 0.5. We report the p-values ranging between 0.0027 - 0.05, and just give the upper limit for highly significant correlations.

The optical part of the plot (upper panel of Figure~\ref{fig:FWHM_Fe}) shows an apparent anti-correlation between the H$\beta$ line width and the FeII equivalent width. 
The trend correlates well with the trend in the black hole mass (see colors of points in Figure~\ref{fig:FWHM_Fe} (upper panel)). 
The dependence does not seem strictly linear in the logarithmic plot, rather suggesting some saturation at values of $\sim 80$ \AA~ when the H$\beta$ line becomes narrower.
Pearson correlation coefficient  (see Table~\ref{tab:coef_opt}) is relatively small (r = -0.39, p-value $<$ 0.0027), which implies that the linear fit does not represent well the data.


However, the UV plot looks visually different (lower panel in Figure~\ref{fig:FWHM_Fe}).
The trend is negative in the optical plane but marginally positive in the UV plane. However, in both cases the correlation is weak (r = -0.39 and 0.13, respectively), so the conclusion is not firm. On the other hand, our determination of the correlation coefficient may not be accurate in the UV since the QSFit catalog contains results obtained apparently with the adoption of the firm lower and upper limit for the EW(FeII), of about 90 \AA~ and 220 \AA, respectively, and we cannot estimate the role of this potential outlier removal on our results. However, in our subsample, the objects do not pile strongly at these limits.
The different behaviour of the FeII emission in the optical and in the UV was already stressed by \citet{kovacevic15} on the basis of the subsample of 293 objects from the SDSS.
They additionally noticed significant redshifts in the FeII$_{UV}$ not seen in the FeII$_{opt}$.

\begin{figure}
 \plotone{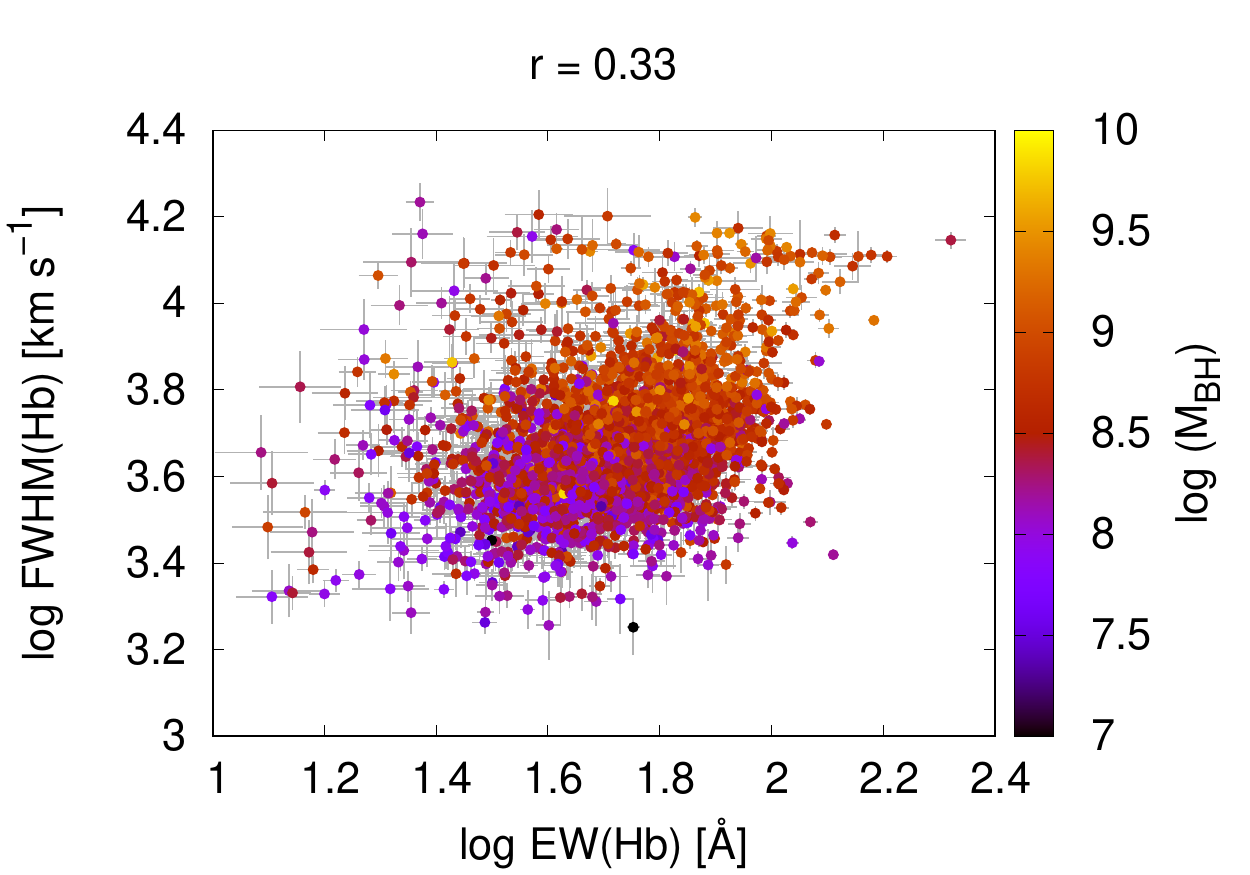}
 \plotone{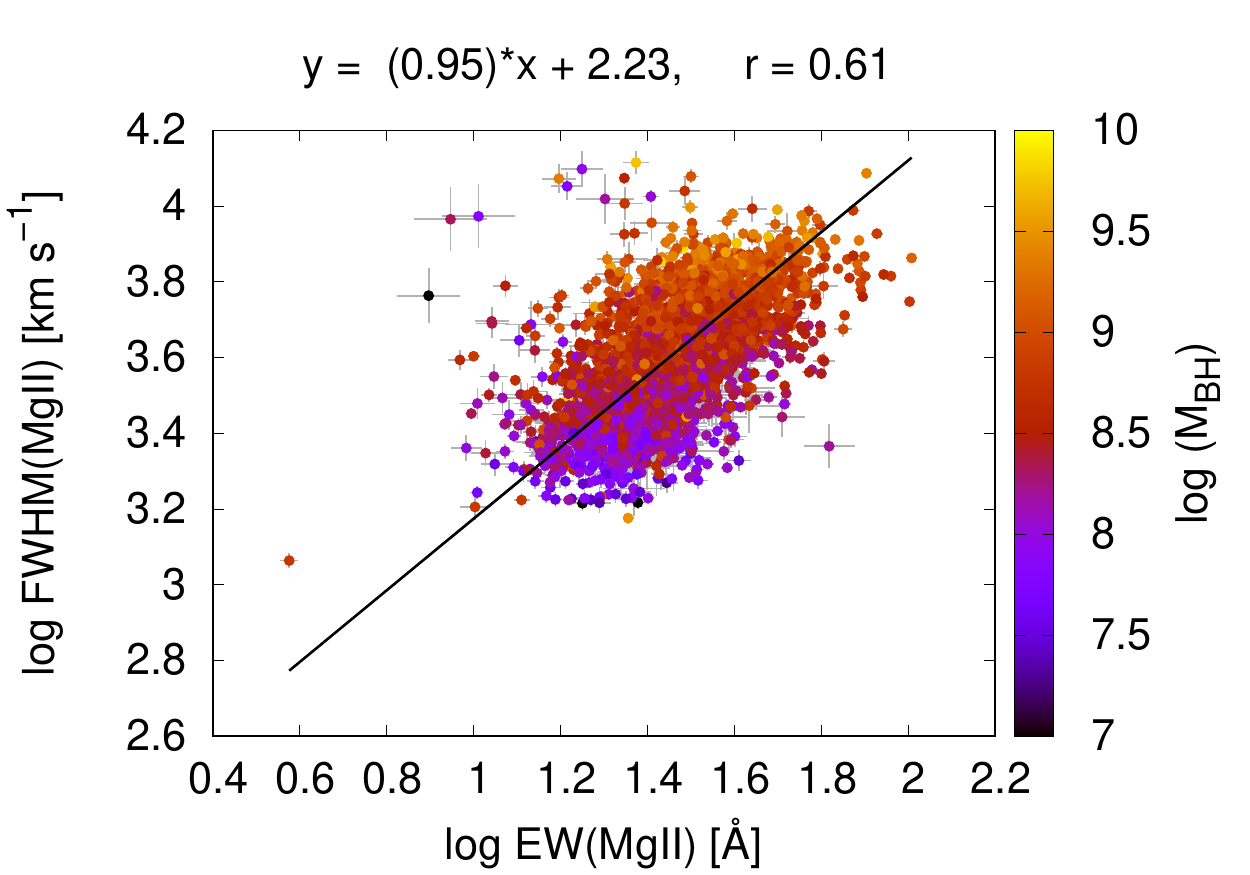}
  \caption{The dependence of the FWHM(H$\beta$) on the EW(H$\beta$) (top), and the FWHM(MgII) on the EW(MgII) (bottom) for the subsample of quasars from the QSFit.  Line fitted using ODR method.}
    \label{fig:FWHM_EW}
  \end{figure}

Next we plot the dependence of the line widths against the corresponding line EWs, again in the logarithmic scale (see Figure~\ref{fig:FWHM_EW}). 

\begin{table*}[]
\caption{The Pearson correlation coefficients and the correlation significances in the optical plane.}   
  \label{tab:coef_opt} 
\centering
\begin{tabular}{lr|rrrr}
    \hline\hline      

                      &   & $\log$ EW(H$\beta$) & $\log$ FWHM(H$\beta$) & $\log$ EW(Fe$_{opt}$) & EW(H$\beta$) \\ \hline
$\log$ EW(H$\beta$)   & r & 1                   & 0.33                  & -0.22                 & $\cdots$     \\
                      & P & 0                   & $<$ 0.0027              & $<$ 0.0027              & $\cdots$     \\ \hline
$\log$ FWHM(H$\beta$) & r & 0.33                & 1                     & -0.39                 & 0.36         \\
                      & P & $<$ 0.0027            & 0                     & $<$ 0.0027              & $<$ 0.0027     \\ \hline
$\log$ EW(Fe$_{opt}$) & r & -0.22               & -0.39                 & 1                     & $\cdots$     \\
                      & P & $<$ 0.0027            & $<$ 0.0027              & 0                     & $\cdots$     \\ \hline
EW(H$\beta$)          & r & $\cdots$            & 0.36                  & $\cdots$              & 1            \\
                      & P & $\cdots$            & $<$ 0.0027              & $\cdots$              & 0            \\ \hline
\end{tabular}

\end{table*}

\begin{table*}[]
\caption{The Pearson correlation coefficients and the correlation significances in the UV plane.}
\label{tab:coef_UV}
                       
\centering 
\begin{tabular}{lr|rrrr}
    \hline\hline      

&   & $\log$ EW(MgII) & $\log$ FWHM(MgII) & $\log$ EW(Fe$_{UV}$) & EW(MgII) \\ \hline
$\log$ EW(MgII)           & r & 1               & 0.61              & 0.38                 & $\cdots$ \\
                          & P & 0               & $<$ 0.0027          & $<$ 0.0027             & $\cdots$ \\ \hline
$\log$ FWHM(MgII)         & r & 0.61            & 1                 & 0.13                 & 0.59     \\
                          & P & $<$ 0.0027        & 0                 & $<$ 0.0027             & $<$ 0.0027 \\ \hline
$\log$ EW(Fe$_{UV}$)      & r & 0.38            & 0.13              & 1                    & $\cdots$ \\
                          & P & $<$ 0.0027        & $<$ 0.0027          & 0                    & $\cdots$ \\ \hline
EW(MgII)                  & r & $\cdots$        & 0.59              & $\cdots$             & 1        \\
                          & P & $\cdots$        & $<$ 0.0027          & $\cdots$             & 0        \\ \hline 
\end{tabular}

\end{table*}

\begin{figure}
 \plotone{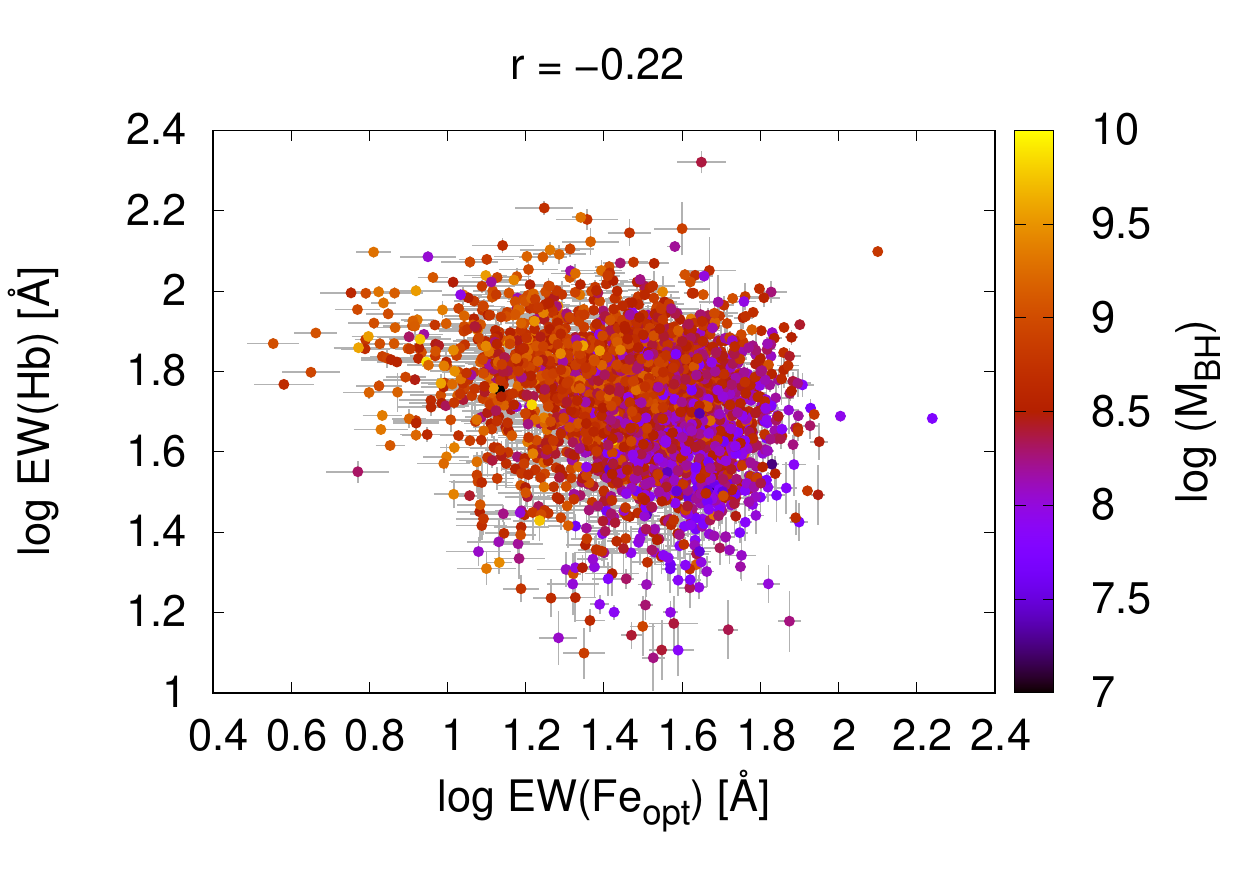}
 \plotone{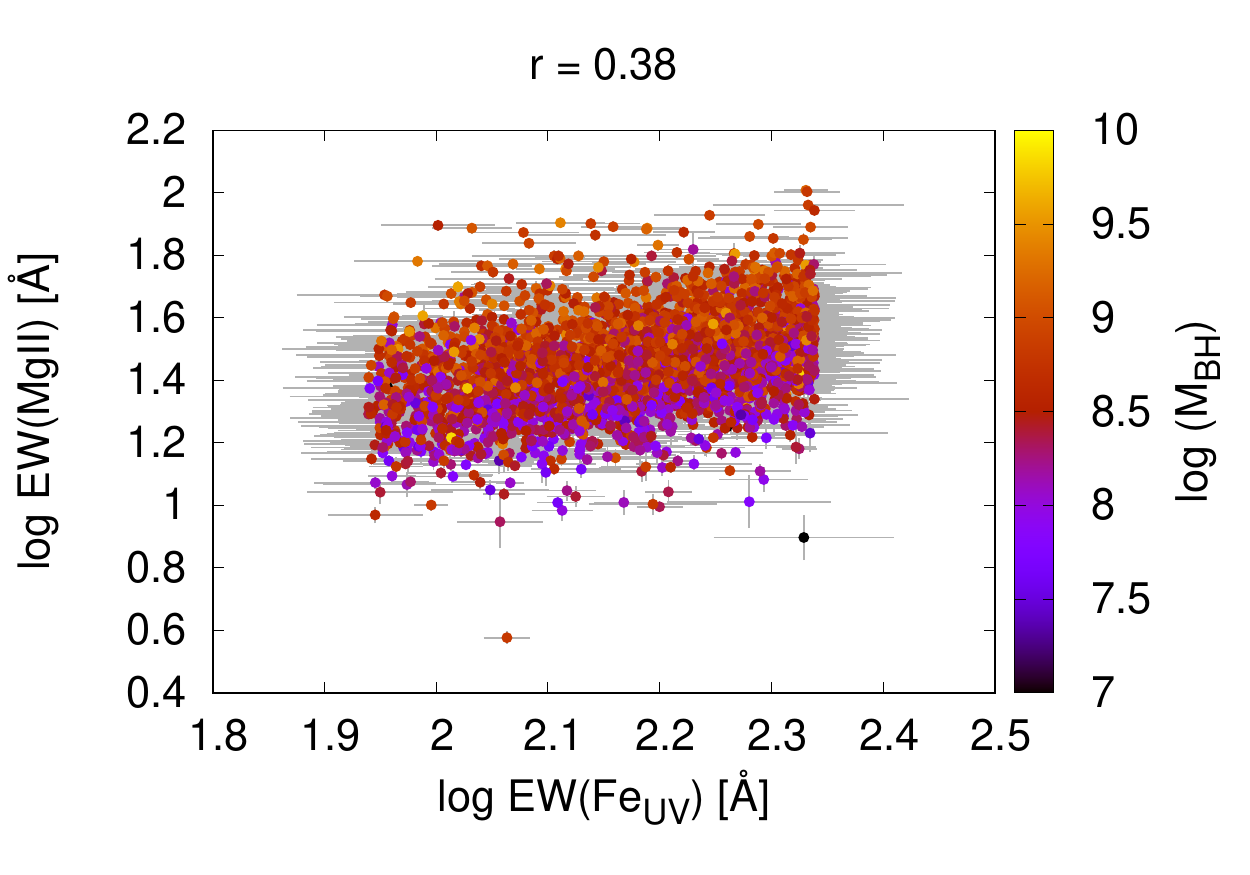}
  \caption{The dependence of the EW(H$\beta$) on the EW(FeII$_{opt}$) (top), and the EW(MgII) on the EW(FeII$_{UV}$) (bottom) for the subsample of quasars from the QSFit.} 
    \label{fig:FWHM_EW_fe}
  \end{figure}

The general rising trend in the FWHM(H$\beta$) with the rising equivalent width is visible (upper panel), apparently correlated with the black hole mass. The correlation is not very strong (r = 0.33) but suggestive.
The correlation is consistent with the trend noticed already by
\citet{osterbrock85} that objects with narrower lines, like the NLS1
galaxies, with line widths smaller than 2000 km s$^{-1}$ have lower line
EWs.
In the case of MgII line (lower panel), 
the correlation between the line width and the line strength itself is very clear as suggested by the high Pearson correlation coefficient, i.e. 0.61 (see Table~\ref{tab:coef_UV}). The slope of the relation is  $0.95 \pm 0.016$, 
consistent with the linear dependence between the two quantities.
On the bottom panel of Figure~\ref{fig:FWHM_EW}, the measurement of the outlying point with the smallest EW(MgII) is affected by absorption present in this particular spectrum. This outlier can also be seen in the lower panel of Figure \ref{fig:FWHM_EW_fe}.
The correlation between the FWHM and the EW of the MgII line has been 
noticed before by \citet{puchnarewicz97} but the strictly linear dependence is only well visible in our relatively larger sample. 

 We tested whether the results for H$\beta$ line are closer to those for MgII line if another method is used to establish the linear trends. We chose the relation FWHM(H$\beta$) vs. EW(H$\beta$), since it showed small values of the coefficient r in the log-log space while the analogous correlation in UV plane was strong. Apart from the orthogonal regression  method with individual errors used throughout the paper we applied several other methods for the best fit linear trend determination \citep{sen, 1992ApJ...397...55F}, listed in Table~\ref{tab:inne} for FWHM(H$\beta$) vs. EW(H$\beta$) and Table~\ref{tab:inne_mg} for FWHM(MgII) vs. EW(MgII). These results do not include the individual measurement errors. Without errors, in the case of H$\beta$, the orthogonal regression returns the coefficient $b$ equal 0.94, instead of 0.83, 
 when the latter is with errors included.

After applying sigma-clipping method for ODR best fits, our sample reduces from 2962 (and $b$ = 0.83) to 105 (and $b$ = 1.14), which simply confirmed that the dispersion in the measurements is much larger than measurement errors, and correlation to too weak to obtain meaningful constraint for the slope.
The standard least squared method gave us much shallower slope (0.32). Similar slopes were obtained by minimum absolute deviation and the Theil-Sen estimator \citep[see][]{FB2012book}. Ordinary Least Square (OLS) bisector and the geometrical mean of the OLS gave much steeper slopes.


So the results are method-dependent, and the slope is determined with relatively very high uncertainty, as expected in the case of a very weak correlation. We repeated similar analysis for the Mg II line, and in this case the slope of the linear fit determined using various methods shows much smaller dispersion.
Thus the trends observed in the optical and UV planes are different.

\begin{table*}[]
\centering
\caption{ Different methods of computation of the least squares fit to y = bx + a, using robust fitting techniques for the log FWHM(H$\beta$) vs log EW(H$\beta$) relation.}
\begin{tabular}{l|ll}
\hline\hline 
log FWHM(H$\beta$) vs log EW(H$\beta$)                                                                        & coefficient b & coefficient a \\ \hline
Standard least squares fit                                                                                & 0.32          & 3.13          \\
Orthogonal distance regression \footnote{Used in this work for all plots, including errors.} & 0.83          & 2.20          \\
Orthogonal distance regression                                                                                & 0.94          & 2.05          \\
OLS bisector                                                                                                  & 0.99          & 1.97          \\
Geometric mean of the OLS lines                                                                           & 0.98          & 1.98          \\
Theil-Sen estimator                                                                                       & 0.35          & 3.07          \\
Minimum absolute deviations                                                                                   & 0.34          & 3.07          \\ \hline
\end{tabular}
\label{tab:inne}

\end{table*}

\begin{table*}[]
\centering
\caption{Different methods of computation of the least squares fit to y = bx + a, using robust fitting techniques for the log FWHM(MgII) vs log EW(MgII) relation.}
\begin{tabular}{l|ll}
\hline\hline 
log FWHM(MgII) vs log EW(MgII)                                                                               & coefficient b & coefficient a \\ \hline
Standard least squares fit                                                                               & 0.65          & 2.66          \\
Orthogonal distance regression \footnote{Used in this work for all plots, including errors}               & 0.95          & 2.23          \\
Orthogonal distance regression                                                                               & 1.10          & 2.00          \\
OLS bisector                                                                                                 & 1.05          & 2.07          \\
Geometric mean of the OLS lines                                                                          & 1.06          & 2.06          \\
Theil-Sen estimator                                                                                      & 0.72          & 2.55          \\
Minimum absolute deviations                                                                                  & 0.74          & 2.52          \\ \hline
\end{tabular}
\label{tab:inne_mg}
\end{table*}

\begin{figure}
 \plotone{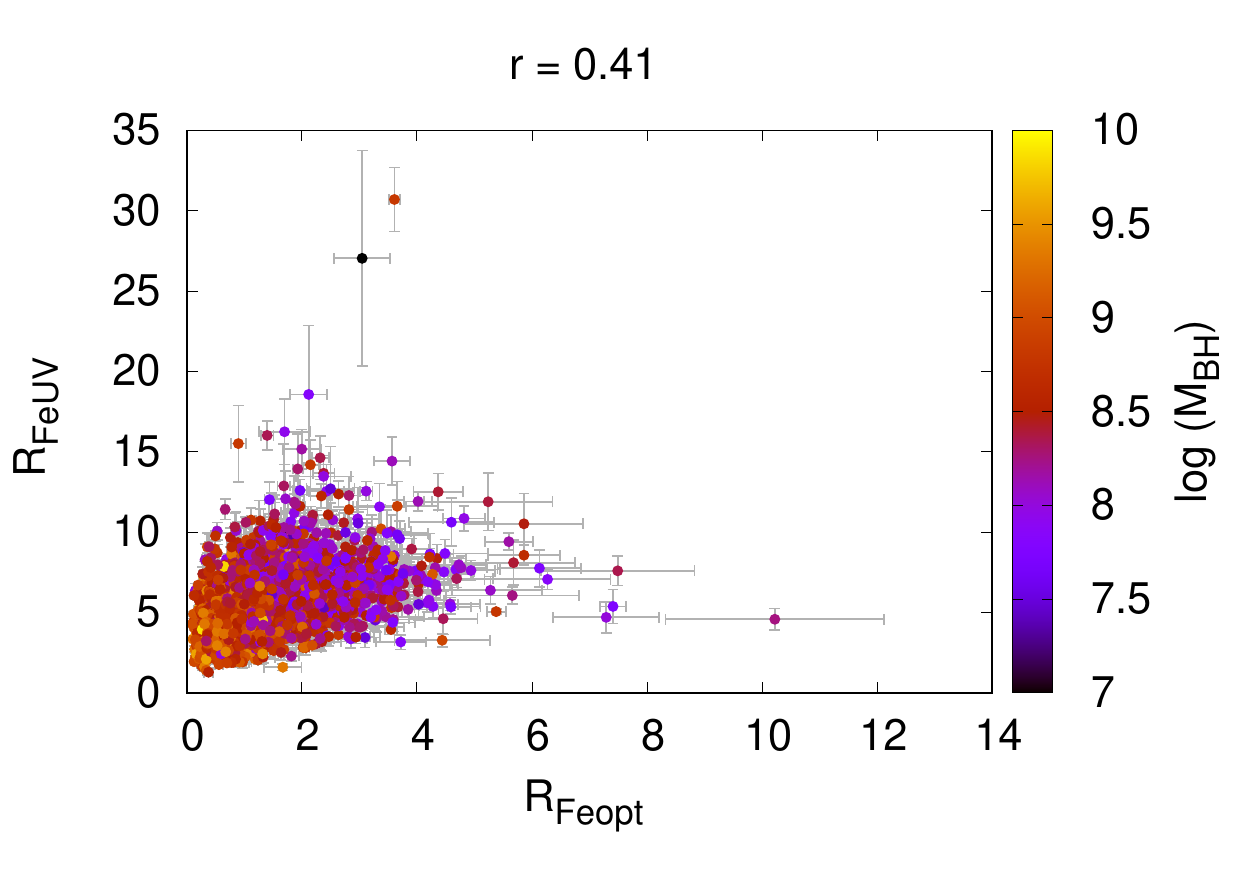}
  \caption{The dependence of the $R_{Fe_{UV}}$ on $R_{Fe_{opt}}$.} 
  \label{fig:rfe_opt_uv}
  \end{figure}

Some apparent similarity of the optical and UV planes in Figure~\ref{fig:planes} and a different behaviour of the corresponding plots based on EWs motivated us to check also those correlations. In Figure~\ref{fig:FWHM_EW_fe} we show the correlation between EW of H$\beta$ or MgII with EW of corresponding FeII emission. The various of the correlations are not very high but they suggest an opposite trend.

Additionally, we checked the correlation between the EW(FeII)  in the optical and UV band and it is very weak (r = 0.042, p-value = 0.0207). The ratios  $R_{Fe_{opt}}$ and $R_{Fe_{UV}}$ 
show a slightly higher correlation, although it is by no means a
strong correlation (r = 0.41, p-value $<$ 0.0027) (see Figure \ref{fig:rfe_opt_uv}).
This is why we also see the UV quasar main sequence in the lower panel of Figure~\ref{fig:planes}, despite no correlations between EWs of the FeII lines in the two bands.


\section{Discussion}
\label{sec:discussion}

We compared the optical and the UV plane of the quasar main sequence. The apparent similarity between the two plots (see Figure~\ref{fig:planes}) is actually misleading, despite the fact that both H$\beta$ and MgII line belong to the same class of Low Ionization Lines, and their widths are plotted there against the corresponding FeII emission. The main sequence in the optical plane is driven both by the trends in FeII and H$\beta$ intensities while in the UV plane the EW(FeII) does not correlate with the MgII line width and the main sequence is based just on the correlation between the MgII line width and intensity.

The tight correlation between EW and FWHM of MgII was first noticed by \citet{puchnarewicz97} in their study of the optical and X-ray properties 
in a sample of
160 X-ray selected AGN from the RIXOS survey.  Similar correlation is observed for the H$\beta$ \citep{osterbrock77,gaskell85,osterbrock85,goodrich89} 
in small samples of objects, but in a large sample of objects such as in the present paper, this correlation is much weaker.

The theoretical interpretation of the observational trends -- strong correlation in the UV and weak correlation in the optical band between the leading line EW and the FWHM -- is not obvious. We first start with common expectations for the optical and the UV emissions based on a specific model of the Broad Line Region, and then discuss the mechanisms which may lead to differentiation between the optical and the UV trends.

\subsection{Comparison of the FWHM and the EW line trends with the BLR model}
\label{sect:FRADO}

Most of the BLR models are parametric, without specific predictions about the line properties. However, Failed Radiatively Accelerated Dusty Outflow (FRADO) model proposed by \citet{ch11} predicts the formation of the emission lines from the LIL group. These lines generally do not show strong asymmetries and thus likely come from the medium with turbulence, but not strong outflow, and the failed wind scenario seems attractive for these lines. We thus compare the model prediction with the observed trend.

Kinematic width of the line (FWHM) is determined by the Keplerian velocity of the material located at a distance $R_{BLR}$ from the central black hole of mass $M_{BH}$

\begin{equation}
 FWHM \propto M_{BH}^{1/2} R_{BLR}^{-1/2}, 
\end{equation}
where we drop constants as well as the geometrical factor related to the dynamics of the BLR and its extension \citep[see e.g.][]{meija18}.

The BLR location in the FRADO model depends only on the monochromatic flux from the disk, usually measured at  5100 \AA. In the model it comes from the fixed effective temperature where the dust forms, and in the observations this is supported by the number of reverberation measurements for (mostly nearby) sources \citep{kaspi00,peterson04,bentz13,dupu15,dupu18}. Thus we have a relationship
\begin{equation}
 R_{BLR} \propto L_{5100}^{1/2},
\end{equation}
and in the FRADO model this slope is exactly 0.5. In the H$\beta$ monitoring \citep[e.g.][]{peterson04} the continuum is measured at 5100 \AA, in the case of MgII monitoring \citep[e.g.][see also \citealt{kozlowski15}]{zhu17,sredzinska17} the continuum is measured at 3000 \AA, which affects only the proportionality coefficient, but not the slope.

From the classical theory of accretion disks \citep{ss73} we know that the spectrum at longer wavelengths is well described by a single power-law shape, with the slope 1/3
\begin{equation}
\label{eq:SS1}
L_{\nu} \propto \nu^{1/3}, 
\end{equation}
where $L_{\nu}$ is the disk luminosity at a frequency $\nu$. The normalization of this power law at a given frequency (or wavelength) depends on the black hole mass, the accretion rate (in dimensional units) and the inclination angle $i$. Thus, for the disk luminosity at 5100 \AA~ we have
\begin{equation}
\label{eq:SS2}
L_{5100} \propto \cos i (M \dot M)^{2/3}.
\label{eq4}
\end{equation}

If we neglect the problem of the viewing angle $i$, which has a limited range of values for the Type 1 objects due to the presence of the dusty-molecular torus, we obtain the following relation of the FWHM on the black hole mass and the accretion rate
\begin{equation}
 FWHM \propto M^{1/3} \dot M^{-1/6}.
\end{equation}

Now we can estimate the line EW as
\begin{equation}
EW \propto {L_{line} \over L_{5100}},  
\end{equation}
where the continuum flux is measured at  5100 \AA\ or 3000 \AA. The line intensity can be estimated roughly as a constant fraction of the incident radiation intercepted by the BLR, and this is linked to the source bolometric luminosity, $L_{bol} = \eta \dot M c^2$, where $\eta$ is the accretion efficiency depending on the black hole spin, and the solid angle  $\Omega_{BLR}$ filled by the BLR   
\begin{equation}
L_{line} \propto \Omega_{BLR} \eta \dot M.
\end{equation}
If we drop the spin-dependent term $\eta$  we obtain a relation 
\begin{equation}
EW \propto  \Omega_{BLR} \dot M^{1/3} M^{-2/3}.
\end{equation}

If the BLR clouds fill the volume densely, the solid angle is given by the geometrical height of the BLR, $z_{max}$, and the distance:
\begin{equation}
  \label{eq:covering_factor1}
\Omega_{BLR} = z_{max}/R_{BLR}
\end{equation}
and combining this relation with the previous one we obtain 
\begin{equation}
EW \propto z_{max} M^{-1}.
\end{equation}
Here we assume that the BLR has axial symmetry, so the solid angle depends linearly on the opening angle.

FRADO model actually predicts the cloud dynamics, so the value of $z_{max}$ as a function of the global parameters of an active nucleus \citep{c15,czerny17}
\begin{equation}
z_{max} \propto \dot M,
\end{equation}
which finally gives
\begin{equation}
EW \propto \dot M M^{-1}.
\end{equation}

The predicted trend is not consistent with the strong trend observed in the UV plane.
In the model, the FWHM should mostly depend on the black hole mass, but the EW should strongly rise with an increase in the accretion rate. This is clearly seen if we express the accretion rate in Eddington (dimensionless) units as $\dot m$ :
\begin{equation}
FWHM \propto M^{1/6} \dot m^{-1/6}; ~~EW \propto \dot m.    
\end{equation}
The dependence on the dimensionless accretion rate in the model implies an expected anti-correlation instead of the observed correlation. Thus, some of the underlying assumptions of the model are apparently incorrect.

The rise of the cloud height with an increase of the Eddington ratio is really generic to the
model, and convenient to explain the line profiles since a larger vertical velocity in proportion to the local Keplerian speed leads to less disk-like profiles. Higher vertical BLR extension is also seen in the data - \citet{kollatschny11} noticed that when modelling the emission line shape with the disk and a turbulent velocity component. 

Therefore, in our opinion a weak point in the derivation above is the assumption that the covering factor of the BLR is close to 1 in the area covered by the clouds, independently from the accretion rate and the vertical extension.
Thus, Eq.~\ref{eq:covering_factor1} in general should be replaced by
\begin{equation}
  \label{eq:covering_factor2}
\Omega_{BLR} = f_c z_{max}/R_{BLR},
\end{equation}
and the covering factor $f_c$ can decrease with the rise in the Eddington ratio. Physically, this is expected when the development of the thermal instability in the rising medium and the cloud formation is taken into account. Such an instability
 is well known for the irradiated media in the context of AGN \citep[see][]{krolik81}. If the material rises high above the disk there is more time for the development of dense compact clouds embedded in a hot fully-ionized plasma out of the initial, moderately dense wind. The quantitative predictions of this phenomenon was not yet done in the context of the FRADO, but it was discussed in a number of papers \citep[e.g.][]{rozanska00,czerny_ch09,rozanska14,rozanska17}. The linear relation between the FWHM and the EW would require
\begin{equation}
f_c \propto M^{1/6} \dot m^{-7/6},
\end{equation}
i.e. a rapid decrease of the covering factor of the BLR with an increase in the Eddington ratio and an increase of the BLR vertical extension. This indeed can be expected if the integrated vertical optical surface density of the BLR zone is roughly constant, independent of the Eddington rate, as actually postulated in \cite{c15}, and the rising clumpiness accounts for the drop in the covering factor and an actual increase in the local cloud density for high Eddington ratio sources, as discussed by \citet{adhikari16,adhikari18}. 

However, this still leaves an important question why the same relation is not seen between the FWHM and the EW for the H$\beta$ while in the (simplified) model predictions both lines should show the same trend. We also analysed the emissivity profiles of the H$\beta$ and the Mg II lines within the BLR clouds, and the properties of the two lines seemed comparable, with no clear systematic differences (see Appendix A).

The explanation might be in the theoretical aspect of the trend predictions. The expected trends between the different parameters characterizing the BLR were easy to formulate analytically since we used an asymptotic power-law with a fixed slope (1/3) (see Equation~\ref{eq:SS1}) and normalization depending on the black hole mass and the accretion rate (see Equation~\ref{eq:SS2}). However, the actual shape of the continuum is not that of an asymptotic power-law, since the disk has the maximum temperature reached in the innermost part which causes the continuum to curve down, and the difference between the asymptotic power-law and the actual continuum systematically increases with the decrease in the wavelength. Since the MgII is located at considerably shorter wavelength, the simplified prediction roughly appropriate for H$\beta$ and the 5100 \AA~band may not apply to the MgII and the 3000 \AA~band. In order to test that, in the following section we modify the approach to the observational data.

\subsection{Effects of the choice of continuum: the observed power-law vs. the asymptotic power-law}
\label{sec:asymptotic}
As stressed by  \citet{goodrich89}, the EW is essentially a combination of the two physical parameters, the line and the continuum luminosity. The two quantities may be physically unrelated since the continuum measured close to the line is not necessarily the driving continuum. This is the case for both the H$\beta$ and the MgII. The H$\beta$ is basically driven by the photons close to 1 Rydberg, creating highly excited hydrogen atoms. Formation of the FeII is even more complex. Therefore a question arises again -- which among these two parameters is driving the correlation between line width and the line EW in the case of the MgII line. The issue of the
underlying power law is even more important for the MgII line since the AGN continuum in the UV is a subject to possible significant extinction, and also, for massive black holes, the true continuum, represented by the underlying accretion disk can start to bend in this region, showing a departure from the power-law continuum. Such a power-law extrapolation has been used for example in the previous section to derive the trends expected from the FRADO model.  

\begin{figure}
 \plotone{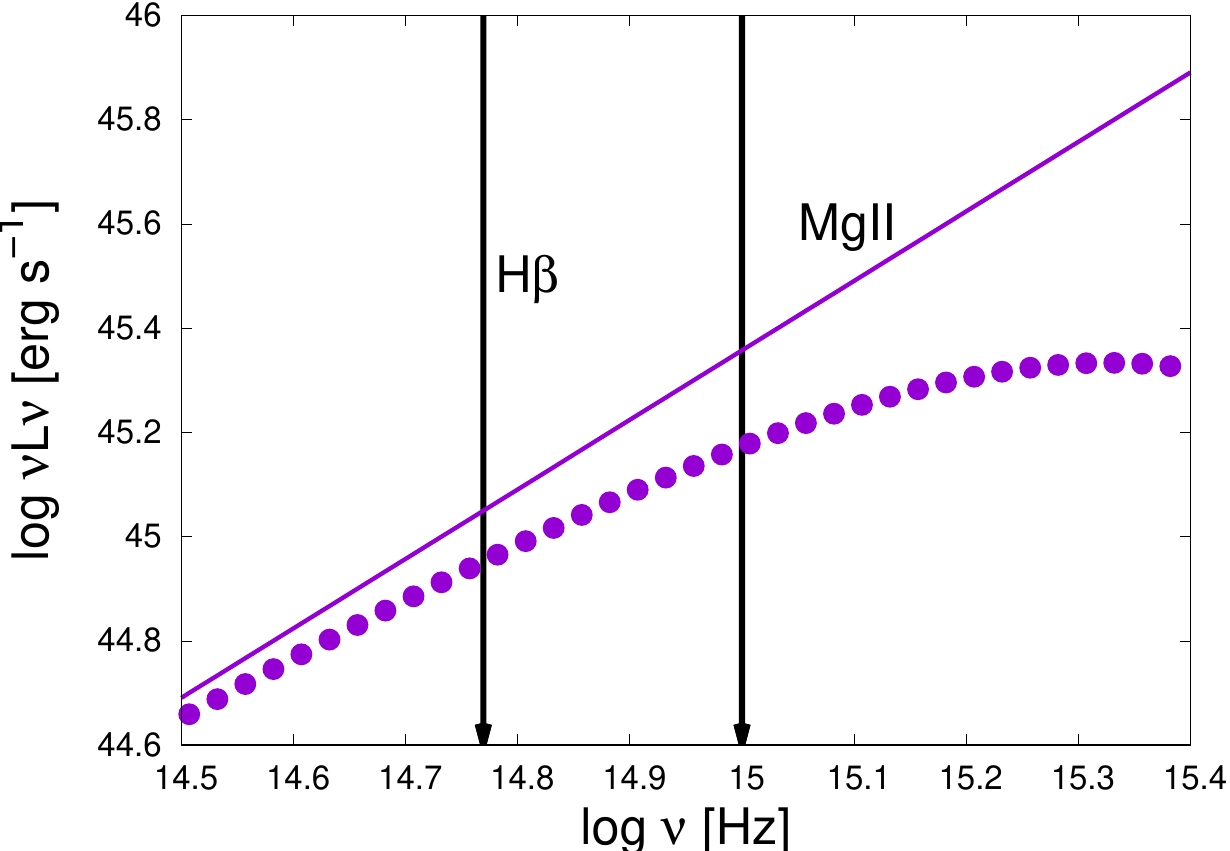}
  \caption{The shape of the disk continuum for a Shakura-Sunyaev disk for a black hole mass $3 \times 10^8 M_{\odot}$ and the Eddington ratio 0.1 (dotted line), and the asymptotic power-law approximation given in Equations \ref{eq:SS1} and \ref{eq:SS2} (continuous line).} \label{fig:spec3e8}
  \end{figure}

The curvature of the predicted continuum shape is noticeable for larger black hole masses and lower accretion rates. In Figure~\ref{fig:spec3e8} we show a disk spectrum and an asymptotic power-law for the average black hole mass in our sample. We see a strong departure from the asymptotic power-law in the MgII line region. For objects with a higher mass and/or lower Eddington ratios the effect is stronger, while in objects with lower masses and/or higher Eddington ratios the asymptotic power-law describes the disk spectrum relatively well. The position of the maximum in the disk spectrum depends on the ratio of $\dot m/M$ for a non-rotating black hole considered here \citep{ss73}. 

We thus perform the following exercise: we calculate a family of accretion disk models using the simplest theory of \citet{ss73} for a range of black hole masses and accretion rates in our sample, and for each model we determine the ratio of the continuum measured at 5100 \AA, and at 3000 \AA~ to the power-law approximation specified by Equations ~\ref{eq:SS1} and \ref{eq:SS2}. Next, for all 2962 objects in our sample we interpolate the results from the grid of the models and we calculate the EWs of H$\beta$, MgII and iron in the optical and the UV range with respect to the power-law instead of the actual continuum by multiplying the EW from the QSFit catalog by the ratio of the disk spectrum to the power-law approximation, appropriate for each object, according to its mass and Eddington ratio reported by the QSFit catalog.
We find a relatively good analytical approximation for this factor as a function of the black hole mass and the Eddington ratio which allows us to apply it conveniently to the whole sample.

\begin{figure}
 \plotone{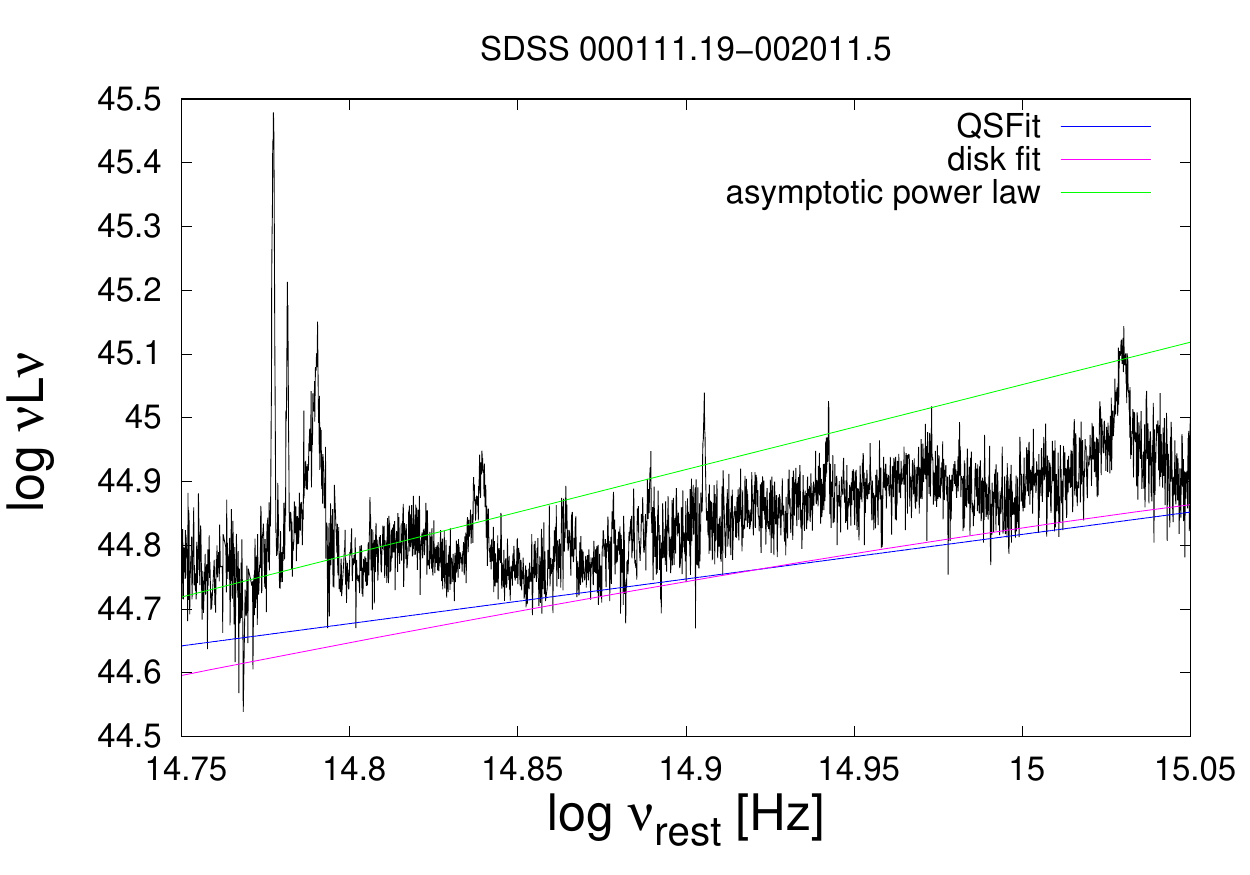}
  \caption{Exemplary quasar spectrum with three continua discussed in this section: best fit continuum from the QSFit (blue line), fit by the Shakura-Sunyaev accretion disk (magenta line) and an asymptotic continuum from the Shakura-Sunyaev model with a fixed slope of $\nu^{1/3}$ (green line). The QSFit represents best the continuum, but the discussion in Sect.~\ref{sect:FRADO} uses the asymptotic continuum.} \label{fig:rec}
  \end{figure}

We call this procedure an exercise since this choice of a new continuum does not imply that the new continuum is a better representation of the observed continuum in the source. We illustrate that in Figure~\ref{fig:rec} for one of the objects SDSS 000111.19-002011.5. Blue line taken from the QSFit catalog - a power-law with the adjusted slope fits the source continuum in the best possible way. The fit with the use of the Shakura-Sunyaev disk model and a black hole mass appropriate for this object is comparable, and it shows some curvature in the spectrum. Stronger effect is expected for a larger black hole mass and a lower Eddington ratio. However, the power-law which we use in this section is an asymptotic power-law, with a fixed slope of $\nu^{1/3}$. It passes well above the whole spectrum, and our only motivation to use it here is purely theoretical. The model predictions for the line behaviour discussed in Section~\ref{sect:FRADO} were based on this asymptotic power-law, without the spectral curvature effect included. Such approach  has lead to the expected similar behaviour of the H$\beta$ and the MgII, while the data shows clear differences in the observed properties of the two lines. The difference between the line behaviour could reflect important differences between the physical properties of the regions where the two lines form, but may be simply related to the lower number of photons at 3000 \AA~continuum than expected from an asymptotic power-law model. It is thus important to stress that by using the asymptotic power-law we do not test the actual curvature of the spectrum but the fact that real spectrum is much redder than suggested by the slope of the asymptotic power-law. 
Thus the accretion disk fit (curved spectrum) and the QSFit continuum (power-law much redder than the asymptotic power-law) are similar, and we use them in the computations performed in this section as equivalent, while the asymptotic power-law is essentially different, and its adoption leads to strongly different values of the EW (see Table~\ref{tab:add}).

\begin{figure*}
\centering
    \includegraphics[scale=0.76]{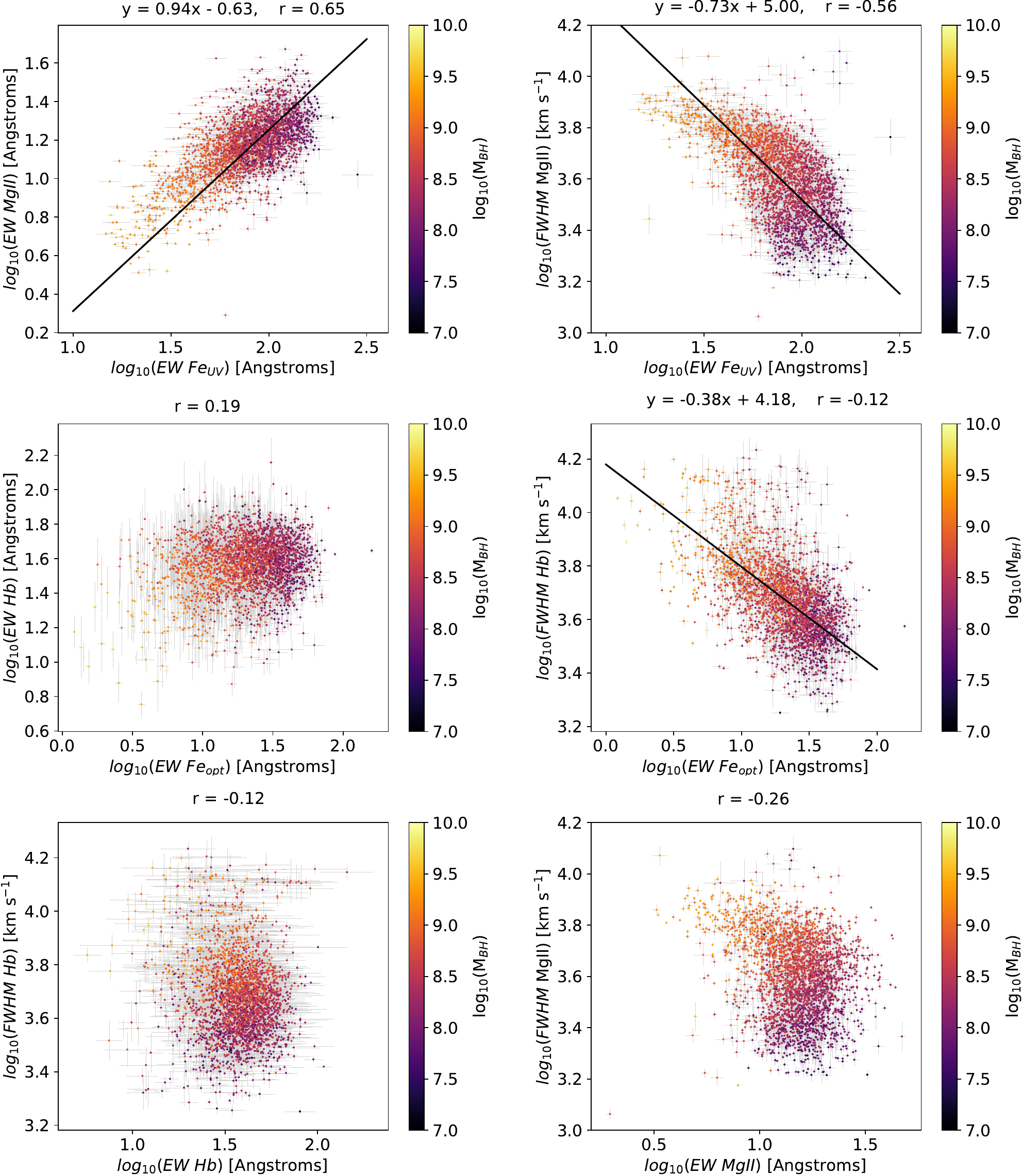}
    \caption{The dependence (from the top left) of the
    EW(MgII) on the EW(Fe$_{UV}$), the FWHM(MgII) on the EW(Fe$_{UV}$),
    the EW(H$\beta$) on the EW(FeII$_{opt}$), the FWHM(H$\beta$) on the EW(FeII$_{opt}$),  the FWHM(H$\beta$) on the EW(H$\beta$) and the FWHM(MgII) on the EW(MgII), when the line EW is measured with respect to the asymptotic power-law. 
    Best fits are shown using ODR method.}
    \label{fig:all_theoretical}
\end{figure*}

In this way we plot new FWHM -- EW relations in the optical and in the UV plane with color coding with the black hole mass (see Figure~\ref{fig:all_theoretical}). 
In tables \ref{tab:coef_opt_2962_5100} and \ref{tab:coef_UV_2962_3000} we show all correlation coefficients for the optical and the UV planes with the EWs measured using the asymptotic power-law.
We see that in this case the relation between the line width and the line EW in log-log scale for the H$\beta$ shows again a weak linear correlation, the Pearson correlation  coefficient is now −0.12, which implies an even weaker correlation than before (r = 0.33). However, for the MgII line the change is substantial, in the case of FWHM(MgII) the strong positive correlation (Pearson coefficient $0.61$) is replaced with a very weak and negative correlation (Pearson coefficient $-0.24$).
Significant change in correlations is also present in the case of iron (for both the optical and the UV). 
 Particular changes are seen in the following correlations: FWHM(H$\beta$) -- EW(FeII$_{opt}$) changed from -0.39 to -0.56, FWHM(MgII) -- EW(FeII$_{UV}$) changed from 0.13 to -0.60 and EW(MgII) -- EW(FeII$_{UV}$) from 0.38 to 0.65.

\begin{table*}[]
\caption{The Pearson correlation coefficients and the correlation significance in the optical plane with the use of the asymptotic power-law (see Figure~\ref{fig:spec3e8}).}   
  \label{tab:coef_opt_2962_5100} 
\centering
\begin{tabular}{lr|rrrr}
    \hline\hline      

                      &   & $\log$ EW(H$\beta$) & $\log$ FWHM(H$\beta$) & $\log$ EW(Fe$_{opt}$) & EW(H$\beta$) \\ \hline
$\log$ EW(H$\beta$)   & r & 1                   & -0.12                 & 0.19                  & $\cdots$     \\
                      & P & 0                   & $<$ 0.0027              & $<$ 0.0027              & $\cdots$     \\ \hline
$\log$ FWHM(H$\beta$) & r & -0.12               & 1                     & -0.56                 & -0.08        \\
                      & P & $<$ 0.0027            & 0                     & $<$ 0.0027              & $<$ 0.0027     \\ \hline
$\log$ EW(Fe$_{opt}$) & r & 0.19                & -0.56                 & 1                     & $\cdots$     \\
                      & P & $<$ 0.0027            & $<$ 0.0027              & 0                     & $\cdots$     \\ \hline
EW(H$\beta$)          & r & $\cdots$            & -0.08                 & $\cdots$              & 1            \\
                      & P & $\cdots$            & $<$ 0.0027              & $\cdots$              & 0            \\ \hline
\end{tabular}

\end{table*}

\begin{table*}[]
\caption{ The Pearson correlation coefficients and the correlation significance in the UV plane with the use of the asymptotic power-law (see Figure~\ref{fig:spec3e8}).}
                       \label{tab:coef_UV_2962_3000}
                       
\centering 
\begin{tabular}{lr|rrrr}
    \hline\hline      

                     &   & $\log$ EW(MgII) & $\log$ FWHM(MgII) & $\log$ EW(Fe$_{UV}$) & EW(MgII) \\ \hline
$\log$ EW(MgII)      & r & 1               & -0.24            & 0.65                 & $\cdots$ \\
                     & P & 0               & $<$ 0.0027          & $<$ 0.0027             & $\cdots$ \\ \hline
$\log$ FWHM(MgII)    & r & -0.24           & 1                 & -0.60                & -0.20    \\
                     & P & $<$ 0.0027        & 0                 & $<$ 0.0027             & $<$ 0.0027 \\ \hline
$\log$ EW(Fe$_{UV}$) & r & 0.65            & -0.60             & 1                    & $\cdots$ \\
                     & P & $<$ 0.0027        & $<$ 0.0027          & 0                    & $\cdots$ \\ \hline
EW(MgII)             & r & $\cdots$        & -0.20             & $\cdots$             & 1        \\
                     & P & $\cdots$        & $<$ 0.0027          & $\cdots$             & 0        \\ \hline
\end{tabular}

\end{table*}

\begin{table*}
\centering
\caption{Values of the EW of the two lines for the quasar SDSS 000111.19-002011.5 shown in Figure~\ref{fig:rec} for various adopted continua.}
\begin{tabular}{l|llll}
\hline
\hline
         & H$\beta$    & MgII     &  &  \\ 
         &  [\AA]   &   [\AA] & \\\hline
QSFit    & 25.55 & 22.79 &  &  \\
DiskFit   & 28.05 & 21.92 &  &  \\
AsymptoticPl & 20.85 & 13.26 &  & \\ \hline
\end{tabular}
\label{tab:add}
\end{table*}

This result, the differences of EWs in the optical and the UV ranges are surprising and important. As it was stressed long ago by \citet{green96}, it is very important what continuum we use to calculate the line EWs before we draw physically motivated conclusions. In the case of the H$\beta$ and the MgII, the continuum normally used for computations is just a matter of convenience - 5100 \AA~ and 3000 \AA~ are located close to the corresponding measured lines and are relatively free of other lines and pseudo-continua contributions. However, they do not represent the actual driving continuum for the two lines. 

In this case, the computing line EWs with respect to the asymptotic power-law is motivated by the theory - this is how the predicted line correlations in Sect.~\ref{sect:FRADO} were obtained. The artificially constructed trend in the FWHM -- EW for the MgII line is roughly consistent with tha theoretical expectation of a weak anti-correlation. 

\section{Conclusions}
\label{sec:conclusions}
We show that the UV plane of the quasar main sequence based on the MgII line and the FeII$_{UV}$ emission looks apparently similar to the optical plane based on the H$\beta$ and the FeII$_{opt}$. The actual linear correlations behind the linear plots are not strong. However, the optical and the UV planes differ with respect to the trend with the FeII emission and with respect to the EW of the corresponding line.

The optical plane shows a well-known weak correlation between the FWHM of the leading line 
and both the FeII and the EW of H$\beta$ line, while the UV plane shows no coupling to the FeII UV emission and, instead, a strong correlation between the FWHM and the EW of the MgII line. Our analysis shows, however, that this strong trend is entirely caused by the choice of the continuum used to measure the EW of the line. The SED of an accretion disk shows a curvature in the UV, and this systematic decrease in the continuum flux at 3000 \AA~ coupled to the black hole mass and the accretion rate causes the correlation. Standard continuum fitting methods account well for this curvature, for example, by adopting the spectral slope as a free parameter. If we artificially correct for this effect in order to have an insight into the physics of the difference between the optical and the UV quasar plane, and use an asymptotic power-law as a reference for both lines, no strong coupling is predicted between the FWHM and the EW for both Low Ionization Lines.

\acknowledgments

{\it Acknowledgements.}
The project was partially supported by the Polish National Science Centre from projects 2015/\-17/\-B/\-ST9/\-03436 (OPUS 9), 2014/\-15/\-B/\-ST9/\-00093 (OPUS), 2014/\-14/\-A/\-ST9/\-00121 (MAESTRO), 2015/18/E/ST9/00580 and 2016/\-21/\-N/ST9/\-03311. We are grateful to the referee for constructive suggestions. We thank Mary Loli Mart{\'i}nez Aldama for helpful comments to the manuscript.


\vspace{5mm}
\facilities{SDSS}
\software{QSFit \citep{calder17}, CLOUDY \citep{2017RMxAA..53..385F}}




\bibliographystyle{aasjournal}
\bibliography{version4_June}

\section*{Appendix A: Comparing Emissivity profiles of FeII in UV and optical}

\begin{figure*}
\begin{center}
  \includegraphics[scale=0.85]{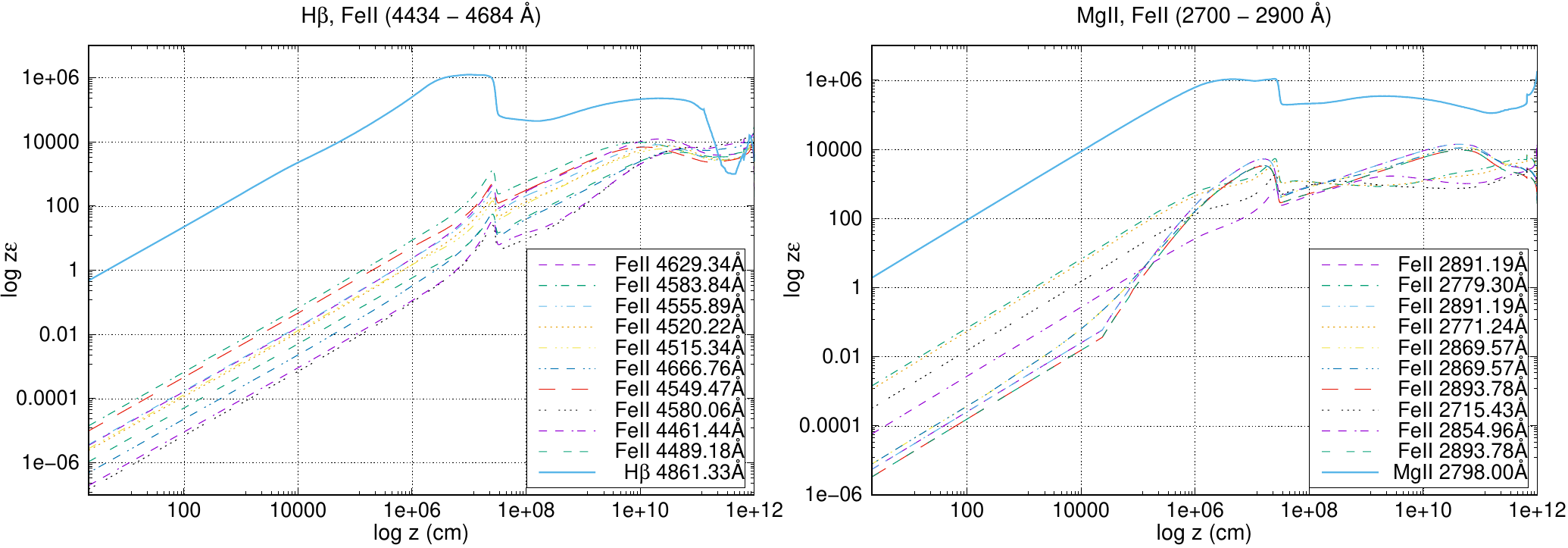}
  \caption{The emissivity profiles of the 10 most intense FeII emission lines in the optical (left panel) and the UV (right panel), with the H$\beta$ and the MgII emission respectively. The profiles shows the emission across the depth of a constant density ($\mathrm{n_{H}}= 10^{12}\;\mathrm{cm^{-3}}$) single BLR cloud.}
  \label{emissivity}
\end{center}
  \end{figure*}

These quasars have the FeII contamination spread throughout the span of their optical-UV spectra. 
Here, we compare the emissivity profiles of the UV FeII (2700-2900 \AA) with respect to the broad MgII emission line (2798 \AA); 
to their optical counterparts (4434-4684 \AA) with respect to the broad H$\beta$ emission line (4861.33 \AA). 
We perform a CLOUDY photoionisation modelling \citep{2017RMxAA..53..385F} assuming a constant density single cloud model. 
It is assumed that all the emission that is considered here (i.e. the optical-UV FeII, the MgII and the H$\beta$) come from the broad line region. 
The spectral energy distribution (SED) of this BLR cloud is modelled by a two-power law based on the prescription given in
\citep{2017FrASS...4...33P,panda18b,2019arXiv190102962P}.
We use the average properties derived from our catalog of 2962 objects that are listed in Table \ref{emtable}. 
The distance to the face of the BLR cloud is derived using the $\mathrm{R_{BLR}-L_{5100}}$ relation using the coefficients from the \textit{Clean}
model of \citet{bentz13}. The cloud's mean density ($\mathrm{n_{H}}$) is assumed to be $10^{12}\;\mathrm{cm^{-3}}$. 
The depth of the cloud is constrained using the \textit{stop column density} command in CLOUDY, which is kept at $10^{24}\;\mathrm{cm^{-2}}$. 
The emissivity profiles are shown in Figure~\ref{emissivity}.

As we can see in the profiles, both the MgII and the H$\beta$ emissions
come from a significant fraction of the volume of
the BLR cloud. At the face of the cloud, the H$\beta$ has at least 4 
orders of magnitude fainter emission
as compared to the optical FeII. This difference in the emission is a factor of 100 lower in the case of the MgII and the FeII in the UV.
Nevertheless, both the profiles show a similar trend with increasing depth, with a peak in the emission (for both H$\beta$ and MgII) showing 
at $\sim$ $10^7$ cm from the face of the cloud. While there is a slump in both these emission profiles after this peak, 
the MgII is shown to recover at around $10^{12}$ cm i.e. the dark edge of the cloud. Throughout the cloud's depth, both the optical
and the UV FeII emission show almost similar behaviour. As a preliminary conclusion, it can be said that both the MgII and the H$\beta$ have extended emitting 
region covering almost the bulk of the BLR cloud, yet the MgII seems to span over a larger region than the H$\beta$.

\begin{table*}
\caption{Catalogue statistics used to produce emissivity profiles}
\label{emtable}
\begin{tabular}{c|cccccc}
\hline
        & \multicolumn{3}{c}{Intrinsic}                                                            & \multicolumn{3}{c}{Derived}                                                  \\ \hline
        & log L$_{bol}$ ($erg\; s^{-1}$) & log L$_{5100}$ ($erg\; s^{-1}$) & log M$_{BH}$ ($M_{\odot}$) & log L$_{UV}$\footnote{at 2500 \AA} ($erg\; s^{-1}$) & log L$_{X}$\footnote{at 2keV} ($erg\; s^{-1}$) & log R$_{in}$ (cm) \\ \hline
        Minimum & 45.055                       & 44.088                        & 7.4                        & 44.107                      & 37.435                     & 17.016            \\
Maximum & 47.16                        & 46.26                         & 9.83                       & 45.955                      & 38.899                     & 18.193            \\
Mean    & 45.852$\pm$0.314             & 44.879$\pm$0.311              & 8.537$\pm$0.371            & 44.786$\pm$0.408            & 38.040$\pm$0.394           & 17.445$\pm$0.169 \\ \hline
\end{tabular}
\end{table*}

\end{document}